\documentclass[aps,pra,reprint,superscriptaddress,amsmath,amssymb,nofootinbib,floatfix]{revtex4-2}
\usepackage{graphicx}
\usepackage{bm}
\usepackage{physics}
\usepackage{braket}
\usepackage[dvipsnames]{xcolor}
\usepackage[hidelinks]{hyperref}
\hypersetup{pdftitle={Memristive associative memories and superposed inputs: A measure-and-prepare limit and a decoder beyond the network},pdfauthor={Karol Bartkiewicz, Patrycja Tulewicz, Adam Miranowicz}}

\begin{document}\title{Memristive associative memories and superposed inputs: A measure-and-prepare limit and a decoder beyond the network}
\author{Karol Bartkiewicz}
\email{bark@amu.edu.pl}
\affiliation{Institute of Spintronics and Quantum Information, Faculty of Physics and Astronomy, Adam Mickiewicz University, 61-614 Pozna\'n, Poland}
\author{Patrycja Tulewicz}
\affiliation{Institute of Spintronics and Quantum Information, Faculty of Physics and Astronomy, Adam Mickiewicz University, 61-614 Pozna\'n, Poland}
\affiliation{Poznan Supercomputing and Networking Center, Institute of Bioorganic Chemistry of the Polish Academy of Sciences, 61-704 Pozna\'n, Poland}
\author{Adam Miranowicz}
\affiliation{Institute of Spintronics and Quantum Information, Faculty of Physics and Astronomy, Adam Mickiewicz University, 61-614 Pozna\'n, Poland}
\date{\today}

\begin{abstract}
The photonic memristor considered here is a single-photon polarization device whose setting depends on the photons detected earlier. In this paper, we ask under which conditions a network of such devices can outperform a classical network used as an associative memory. We first describe how the Hebbian couplings of an attractor memory can be built by sending sequences of photons through pairwise-coupled devices. If the devices forget at a finite rate, this memory works as a palimpsest. Next, we add three-body couplings, which can be provided by a three-photon controlled-controlled-phase gate without optical nonlinearity. Our classical simulations for 64 to 256 spins show that the number of random patterns that are recovered exactly in this case grows faster than linearly with the number of spins, although three-body couplings store no more bits per coupling than two-body ones. We also tested a short linear quantum-annealing schedule for the recall of corrupted classical patterns. Unfortunately, we found no detectable gain for the system sizes simulated, because a zero-temperature descent with the same input bias, set within a window of weak bias strengths, performs equally well at moderate noise. We then turn to a memory task in which the register receives an equal superposition of two stored words, and neither the pair of words nor their relative phase is known. We show that if at least three words are stored, then no measure-and-prepare memory, including any readout that returns a single configuration, can reach a mean fidelity above one half. Unitary evolution under the network couplings does not help here, as the fidelity is about as low as without the couplings. However, the superposition can be restored with unit fidelity by a coherent decoder that writes the error into a syndrome ancilla, provided that the same bit-flip error, within the decoding radius of the stored words, acts on both branches. The drawback of this decoder is that it corrects neither phase errors nor photon loss. It should be stressed that this is a standard decoder that does not use the memristor, and whether the memristive feedback law itself can store such words is not resolved in this paper.
\end{abstract}

\maketitle

\section{Introduction}
\label{sec:intro}

Photonic hardware is a leading platform for fast and energy-efficient neuromorphic computing~\cite{Shastri2021}. The elementary nonlinear ingredient of such hardware is the quantum memristor, i.e., a memory-dependent quantum device driven by measurement feedback~\cite{Pfeiffer2016,Salmilehto2017,Sanz2018}. Photonic implementations now include an integrated-waveguide device~\cite{Spagnolo2022}, neuromorphic computing on a photonic quantum memristor~\cite{Selimovic2025}, vacuum--one-photon qubits from a quantum-dot source~\cite{DiMicco2025}, and a pair of coupled integrated memristors with crossed feedback~\cite{Baldazzi2026}. In Ref.~\cite{BartkiewiczPartI2026}, hereafter Part~I, we analyzed a single-photon memristor with polarization encoding, in which the detection history is stored in the angle of a half-wave plate. We gave there the exponential memory law of its register, and we treated the memory of the feedback device as a classical register. For a coherent extension with delay loops, we also gave closed-form loss budgets for a steering test of entanglement storage.

Associative memories of the Hopfield type~\cite{Hopfield1982} are based on dissipative attractor dynamics. Quantum generalizations of these memories have been studied~\cite{Ventura2000,Trugenberger2001,Rebentrost2018,Rotondo2018,Miller2021,Kimura2025}, including cavity-QED proposals~\cite{Fiorelli2020,Marsh2021} and an experiment~\cite{Marsh2026}. A single driven-dissipative nonlinear oscillator acts as a quantum associative memory~\cite{LabayMora2023} that also stores squeezed states and coherent superpositions~\cite{LabayMora2024}, and a dissipative quantum walk has been shown to act as autonomous bit-flip correction that keeps the coherence between patterns~\cite{LabayMora2025}. However, pattern completion of classical data is a classical task, and a claim of quantum advantage for it needs justification. Hopfield-type storage of entangled states has also been proposed, with a capacity that scales faster than a naive analogy with the classical model suggests~\cite{Hu2026}. 

\begin{figure*}[tb]
\centering
\includegraphics[width=\textwidth]{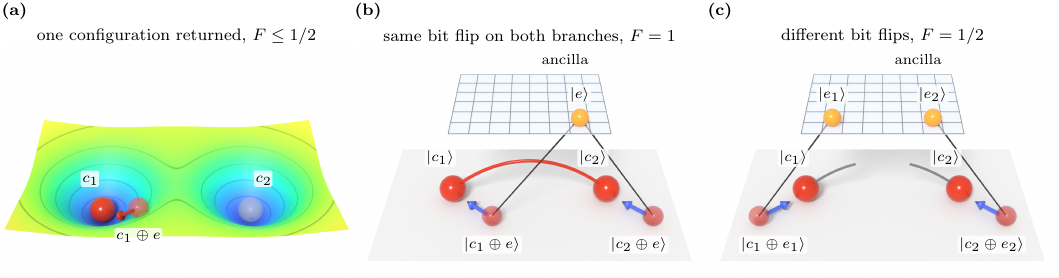}
\caption{Classical and coherent recall (schematic). The surface in (a) is a sketch, which we do not compute here, of an energy surface whose wells are two stored words $c_1$ and $c_2$. (a)~Retrieval takes one corrupted classical input to one well, and the second word is lost (pale sphere). As a result, a readout that returns one configuration has fidelity $F$ (the overlap of the returned state with the target, Sec.~\ref{sec:coherent}) at most $1/2$ with the equal superposition of the two words. (b)~A coherent decoder (a unitary map that does not perform a descent on an energy surface) takes both branches of a superposition to their words (straight arrows above the grey plane, which only marks their positions) and writes the error into an ancilla, shown as a plate whose coordinates label the error. If the same bit-flip error, within the decoding radius (the largest number of bit flips that the decoder corrects, Sec.~\ref{sec:coherent}), acts on both branches, the ancilla holds one state, which does not reveal the word, and the coherence (arc) survives with $F=1$. (c)~If the two branches carry different bit-flip errors, the ancilla records the branch and the coherence is lost (broken arc), with $F=1/2$. The memristive network of Fig.~\ref{fig:network} realizes the retrieval of (a). Note that the decoder of (b) and (c) is a standard one and does not use the memristor.}
\label{fig:concept}
\end{figure*}

In this paper, we study what a network of the devices of Part~I computes and under which conditions it can outperform a classical network. We also consider how a memory recalls an input that is itself superposed. We report a gain only for a superposed input, where it comes from a coherent decoder outside the network, as sketched in Fig.~\ref{fig:concept}, whereas in the retrieval of this network on a linear schedule we find no detectable quantum gain. Each device keeps its memory in a classical register, so storage in it is measure-and-prepare and obeys the limit given below, while the coherent loop applies a unitary to the register alone and therefore cannot contract corrupted inputs onto stored words. A decoder that writes the error into an ancilla has neither of these limitations. We obtained the following results. (i)~We define a memory task in which the register receives an equal superposition of two of $K$ stored words, where the pair and the relative phase are unknown. Such an input cannot be accepted by the retrieval of a memristive Hopfield network. For $K\ge3$ the mean fidelity of every measure-and-prepare memory, including any readout that returns one configuration, is at most $1/2$. This is a measure-and-prepare benchmark of the kind used for quantum storage~\cite{FuchsSasaki2003,Calsamiglia2009}, which is applied here to an unknown pair, and it bounds every memory that stores its input in a classical register. We show that unit fidelity is reached by a coherent decoder with a syndrome ancilla under any bit-flip channel on the register whose errors lie within the decoding radius of the stored words. This is possible since each such error acts alike on both branches. No measure-and-prepare memory can match this tolerance, and in this paper it is the only gain with this property. It should be stressed that the decoder does not correct phase errors or photon loss. Whether the memristive feedback law can store the required words is not resolved in this paper (Sec.~\ref{sec:coherent}). (ii)~Since a single memristor cannot correct the error of the photon it transmits, pattern completion is a phenomenon at the level of the network (Sec.~\ref{sec:device}). (iii)~Hebbian learning~\cite{Hebb1949} is realized in hardware when sequences of photons are streamed through pairwise-coupled memristors, in the mean-field limit of the feedback law for slow forgetting. Already at ten detected coincidences per pair and pattern the correlation of the streamed couplings with the Hebbian matrix is $0.95$ (Sec.~\ref{sec:hebbian}). (iv)~A short quantum-annealing schedule~\cite{Kadowaki1998,Seki2015} improves the recall of corrupted patterns over greedy descent only because the corrupted input enters the annealed problem as a longitudinal bias. Provided that the bias strength $h$ lies in the window $0.2\le h\le0.3$ and the noise is low enough for the network to retrieve, a zero-temperature descent given the same bias matches or exceeds both annealers on registers of $n=8$, $12$ and $14$ qubits (at $n=14$ at one noise level). At the stronger bias $h=0.6$ this holds only at $n=8$. Over the whole range that we can simulate exactly, $n=6$--$14$, the exact-recovery capacity stays at one pattern (Sec.~\ref{sec:retrieval}). (v)~With finite forgetting the network becomes a palimpsest, whose memory window is tuned by the relaxation rate $\gamma$ (Sec.~\ref{sec:palimpsest}). (vi)~Finally, we consider an extension of the architecture of Part~I with delay loops, in which each controlled-phase gate decomposes exactly into an Ising bond and local fields. We propose to upgrade the loop gate to a three-photon controlled-controlled-phase gate, which supplies three-body couplings without a physical optical nonlinearity. Our classical simulations of the resulting energy show that the exact-recovery capacity for random patterns grows superlinearly with the network size between $n=64$ and $256$, and that it passes the $n$ words of an orthogonal code (Sec.~\ref{sec:loop}).

The paper is organized as follows. In Sec.~\ref{sec:device} we recall the device and its single-device limit. The network, its recall of classical patterns, forgetting and a benchmark on optical states are presented in Sec.~\ref{sec:hopfield}. We then define the memory task and treat coherent recall in Sec.~\ref{sec:coherent}, and in Sec.~\ref{sec:loop} we describe the coherent loop and the higher-order couplings. We conclude in Sec.~\ref{sec:conclusions}. The numerical methods are given in the Appendix.

\section{The device and its limit}
\label{sec:device}

\begin{figure}[tb]
\centering
\includegraphics[width=\columnwidth]{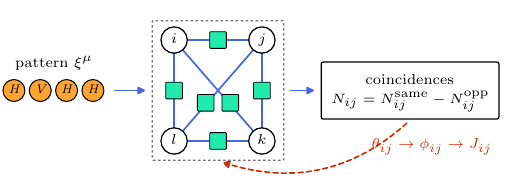}
\caption{Memristive network. Photons encoding a pattern $\xi^\mu$ enter a register in which each pair of qubits has its own memory variable (squares), updated as in Eq.~\eqref{eq:theta_dynamics}. Blue arrows: photons. Blue lines: coupled pairs. Red dashed arrow: signed coincidences update the classical variables $\theta_{ij}$ on all pairs, which program the phases of two-photon gates and through them the couplings $J_{ij}$.}
\label{fig:network}
\end{figure}

We recall from Part~I the minimal description of the device, which implements a memory-dependent unitary transformation
\begin{equation}
\ket{\psi_\text{out}(t)} = \hat{U}(\theta(t))\ket{\psi_\text{in}(t)},
\label{eq:qm_transform}
\end{equation}
realized for polarization qubits by a half-wave plate with its fast axis at angle $\theta$,
\begin{equation}
\hat{U}(\theta) = \begin{pmatrix}
\cos 2\theta & \sin 2\theta \\
\sin 2\theta & -\cos 2\theta
\end{pmatrix},
\label{eq:hwp}
\end{equation}
written in the basis $\{\ket{H},\ket{V}\}$. The memory variable is classical, and it evolves by feedback after each detection together with relaxation,
\begin{equation}
\frac{d\theta}{dt} = \eta\, N(t) - \gamma(\theta - \theta_0),
\label{eq:theta_dynamics}
\end{equation}
where $N(t)$ is the point process of photon detections, $\eta$ is the feedback strength, and $\gamma$ is the relaxation (forgetting) rate toward the rest angle $\theta_0$. The mean register integrates the detection probability with the memory kernel
\begin{equation}
\mathcal{K}(\tau) = \eta\,\Gamma_\text{ph}\,e^{-\gamma\tau},
\label{eq:memory_kernel}
\end{equation}
where $\Gamma_\text{ph}$ is the input photon rate (see Part~I for the derivation and for the steering test of the coherent extension). In Fig.~\ref{fig:network} we show the network studied here.

Note that the node itself provides no error tolerance, although the usefulness of an associative memory lies in completing corrupted inputs. For a fixed $\theta$ the device acts as the unitary of Eq.~\eqref{eq:hwp}, which reflects a linear polarization across the plate axis and preserves inner products. An ideal input $\ket{\psi_0}$ and a perturbed copy $\ket{\tilde{\psi}}$ with fidelity $F_\text{in}=|\!\braket{\psi_0|\tilde{\psi}}|^2$ are thus mapped to outputs with the same fidelity,
\begin{equation}
F_\text{out}=|\!\braket{\hat{U}\psi_0|\hat{U}\tilde{\psi}}|^2=F_\text{in}.
\label{eq:no_denoise}
\end{equation}
For a given photon this is not changed by feedback, since $\theta$ at its passage does not depend on this photon. Feedback can suppress only an error that persists over the response time of the feedback. Consequently, pattern completion must emerge from the attractor dynamics of a coupled register~\cite{Hopfield1982}.

\section{Memristive quantum Hopfield network}
\label{sec:hopfield}

In this section, we build an associative memory from the devices of Part~I. We first show in Sec.~\ref{sec:hebbian} that the feedback law of each device stores the Hebbian couplings. In Sec.~\ref{sec:retrieval} the retrieval of corrupted classical patterns is compared with classical controls, and finally we apply the network to denoising of optical quantum states (Sec.~\ref{sec:qsdl}).

\subsection{Hebbian learning from the feedback law}
\label{sec:hebbian}

\begin{figure*}[tb]
\centering
\includegraphics[width=\textwidth]{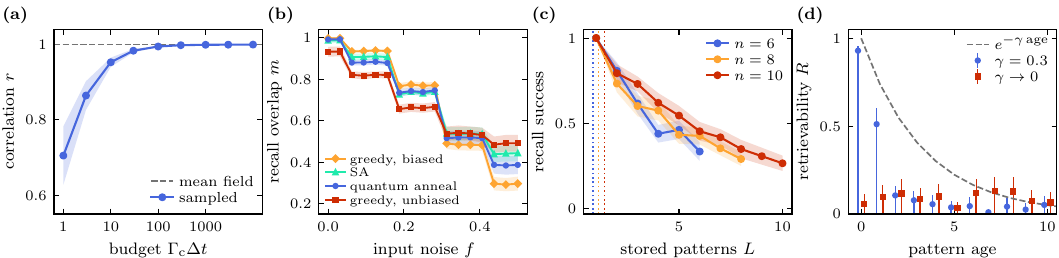}
\caption{Memristive quantum Hopfield network. (a)~Correlation $r$ of the streamed couplings with Eq.~\eqref{eq:hebbian} versus the coincidence budget per pair and pattern ($n=8$, $L=5$, mean $\pm$ standard deviation over $200$ streams). The dashed line is the mean-field limit. (b)~Recall overlap $m$ [Eq.~\eqref{eq:overlap}] versus the input-noise fraction $f$ ($n=8$, $L=3$, $60$ realizations $\times$ $100$ trials per point, shaded $95\%$ confidence intervals across realizations). The quantum anneal, simulated annealing (SA) and greedy descent receive the corrupted input as a bias $h_i=0.6\,\tilde\xi_i$. Greedy descent and SA start from $\tilde\xi$, and the $\Omega=0$ control relaxes on the unbiased couplings from the same configuration. The staircase reflects the discrete number of flipped spins, $\lfloor fn+1/2\rfloor$, which changes at odd multiples of $1/(2n)$. Beyond $f\approx0.3$ the probe leaves the basin of attraction and the ranking depends on the sign-blind definition of $m$, which reverses it. (c)~Recall success of the quantum anneal (fraction of probes recovered exactly, flipped spins $\mathrm{round}(0.125\,n)$) versus the number of stored patterns for $n=6,8,10$ ($60$ networks per point, shaded $95\%$ confidence intervals), with the asymptotic critical load $0.14\,n$ (dotted). (d)~Retrievability, the fraction of probes recovered exactly, versus pattern age, i.e., the number of patterns presented after it, for $\gamma=0.3$ ($60$ streams, $95\%$ confidence intervals), compared with the non-forgetting limit $\gamma\to0$ (squares). The latter lies at a noise floor of about $9\%$ (mean over ages, $6\%$ for the newest pattern) with no age structure. The dashed curve is the kernel weight $e^{-\gamma\,\mathrm{age}}$, shown for reference.}
\label{fig:qhnm}
\end{figure*}

We assign one memristor to each pair of qubits, so that each coupling has its own memory variable $\theta_{ij}$, which is driven through Eq.~\eqref{eq:theta_dynamics} by the signed coincidence signal $N_{ij}(t)=N^{\text{same}}_{ij}(t)-N^{\text{opp}}_{ij}(t)$. Here a coincidence of two photons of the same polarization ($HH$, $VV$) increments $\theta_{ij}$, whereas this variable is decremented if the polarizations are opposite ($HV$, $VH$). The mean drive is then proportional to the correlator $\xi_i\xi_j$ of the pattern values $\xi_i,\xi_j=\pm1$ presented on qubits $i$ and $j$. Since a wave plate on one qubit has no two-qubit action, $\theta_{ij}$ acts on the register only as a classical setting, which is read out to program the phase of a two-photon gate (Sec.~\ref{sec:cz_ising}). If the plate stays in the detected path, the drive acquires a factor $\cos4\theta_{ij}$, which requires $|\theta_{ij}|\ll\pi/8$. By calibrating the phase so that $J_{ij}\propto\theta_{ij}$, we define a network with the transverse-field Ising Hamiltonian of the quantum Hopfield model~\cite{Nishimori1996}
\begin{equation}
\hat{H} = -\sum_{i<j} J_{ij}\,\hat{\sigma}_z^i\hat{\sigma}_z^j - \Omega\sum_i \hat{\sigma}_x^i - \sum_i h_i\,\hat{\sigma}_z^i,
\label{eq:qhn_hamiltonian}
\end{equation}
where $\hat{\sigma}_x^i$ and $\hat{\sigma}_z^i$ are Pauli operators of qubit $i$, the transverse field $\Omega$ enables tunneling between stored configurations and $h_i$ are local biases. For learning we need, on all pairs, only the sign given by which detector pair fired. A register of $n$ qubits stores $L$ patterns $\xi^\mu\in\{+1,-1\}^n$, $\mu=1,\dots,L$, where each pattern is encoded as $\xi_i=+1\mapsto\ket{H}\equiv\ket{0}$ and $\xi_i=-1\mapsto\ket{V}\equiv\ket{1}$ on qubit $i$, i.e., in the eigenstates of $\hat\sigma_z^i$ with eigenvalues $\pm1$. By integrating Eq.~\eqref{eq:theta_dynamics} over one presentation of duration $\Delta t$, we obtain for this pattern the weight $(\eta\Gamma_{\rm c}/\gamma)(1-e^{-\gamma\Delta t})$, where $\Gamma_{\rm c}$ is the coincidence rate. For slow forgetting ($\gamma L\Delta t\ll1$) the memory variables converge to
\begin{equation}
\theta_{ij}\to\eta\,\Gamma_{\rm c}\,\Delta t\sum_\mu\xi_i^\mu\xi_j^\mu,
\label{eq:hebbian_streamed}
\end{equation}
which is, up to an overall scale, the textbook Hebbian matrix~\cite{Hebb1949,Hopfield1982}
\begin{equation}
J^\text{Hebb}_{ij}=\frac{1}{n}\sum_\mu\xi_i^\mu\xi_j^\mu.
\label{eq:hebbian}
\end{equation}
Equation~\eqref{eq:hebbian_streamed} is the mean of the feedback law, and its agreement with the deterministic correlator drive is an algebraic identity. Since Eq.~\eqref{eq:theta_dynamics} is driven by a point process, we have to determine how many detections the identity requires. For this purpose, we sample the signed coincidence counts as independent Poisson variables of mean $\Gamma_{\rm c}\Delta t\,(1\pm\xi_i\xi_j)/2$ and stream $L=5$ patterns through an $n=8$ register [Fig.~\ref{fig:qhnm}(a)]. We measure the quality of the weights accumulated in the registers by the Pearson correlation $r$ of the streamed couplings with Eq.~\eqref{eq:hebbian}. This correlation is set by the coincidence budget per pattern, and it rises from $r=0.71\pm0.08$ at one coincidence per pair and pattern to $r=0.95$ at ten, while beyond $10^{3}$ coincidences we find $r>0.999$. The same coincidences are obtained with attenuated polarized light, and thus these statistics are classical.

\subsection{Retrieval of classical patterns}
\label{sec:retrieval}

For retrieval we freeze the couplings and inject a corrupted pattern $\tilde{\xi}$ with a fraction $f$ of spins flipped as a longitudinal bias. We prepare the register in the ground state $\ket{+}^{\otimes n}$ of the transverse field, with $\ket{+}=(\ket{0}+\ket{1})/\sqrt2$, and evolve it toward the ground state of Eq.~\eqref{eq:qhn_hamiltonian} through a short schedule that lowers $\Omega$~\cite{Kadowaki1998}. We quantify success by the overlap between the retrieved configuration $s\in\{+1,-1\}^n$ and the probed pattern $\xi$,
\begin{equation}
m(\xi,s)=\frac{1}{n}\Big|\sum_{i=1}^{n}\xi_i\,s_i\Big|,
\label{eq:overlap}
\end{equation}
which ignores the overall sign because Hopfield attractors come in pairs $\pm\xi$. The input bias breaks this symmetry and, consequently, $m$ is reliable for the biased rules only while the signed overlap agrees with it, which holds up to $f\approx0.28$.

In Fig.~\ref{fig:qhnm}(b) we show $m$ versus $f$ for $n=8$ and $L=3$, where the transverse field improves recall over greedy $\Omega=0$ descent. However, this comparison favors the anneal, because the annealed problem receives the corrupted input as a longitudinal bias $h_i=h\,\tilde\xi_i$ of strength $h=0.6$ (Appendix), which is the usual encoding of the probe in annealing recall~\cite{Neigovzen2009,Seddiqi2014,Santra2017}. It is noted in these studies that a classical network receives the probe only as its initial state~\cite{Neigovzen2009,Santra2017}, but none of them runs a classical solver on the same biased energy. Our greedy control, on the other hand, relaxes on the unbiased couplings and only starts from $\tilde\xi$. We observe that the effect disappears when the bias is restored to the greedy control, which at this size and bias rules out barrier crossing as its origin. It should be noted, however, that descent receives the input twice, i.e., as the field and as the starting point, and that from a random start it falls below both annealers. For $f=0.125$ the biased zero-temperature descent returns the same configuration as simulated annealing on $95.9\%$ of $6000$ trials, and it exceeds the quantum schedule by $0.054\pm0.016$ in mean overlap. The unbiased control does not reach unit overlap even on uncorrupted inputs (mean overlap $0.93$, exact recovery on $78\%$ of probes at $f=0$), where there is no spurious minimum to escape. We have chosen $L=3$, which is above the exact-recovery capacity of one pattern quoted below. With $L=1$ all three biased rules return the stored pattern exactly, and at $L=2$ and $f=0.125$ they coincide with the unbiased control ($m=0.94$), since $h=0.6$ is then too strong (at $h=0.2$ biased descent reaches $0.98$--$0.99$). Thus, only an overloaded network, in which recall is partial, has spurious minima that a transverse field could help to escape. At $f=0.1875$ the overlap is $0.73$--$0.77$, which should be compared with $0.27$ for a random string and $0.50$ for returning the corrupted input. We find that at this size and bias the biased descent is the best of the four rules up to $f\approx0.28$. At $f=0.1875$ it exceeds simulated annealing by $0.020\pm0.006$ when averaged over six further draws of $60$ networks (the plotted draw, at $0.040\pm0.017$, is high), and the linear quantum anneal never exceeds it in this range. Beyond $f\approx0.3$ the probe has left the basin of attraction and every method has lost most of its recall. Simulated annealing then exceeds descent in $m$ ($+0.043\pm0.004$ at $f=0.3125$ over further draws), but this difference comes from configurations of the wrong sign. If the sign is kept, descent is above simulated annealing ($+0.013\pm0.004$), while the quantum schedule is above both ($+0.018\pm0.003$ over descent).

To exclude an artifact of one register size or one bias strength, we scanned both under the same protocol at fixed $L=3$, so that the load $L/n$ falls with size, with networks and probes shared across bias values. We first cross five noise fractions ($f=0.0625$ to $0.3125$) with five bias values ($h=0.1$ to $0.6$) at $n=8$ and $12$, which makes ten (size, noise) cells with $20$ networks $\times$ $100$ trials per cell. In addition, the bias is varied from $0.1$ to $2.4$ at $f=0.1875$ for $n=8$, $12$ and $14$, while at $n=14$ we ran the noise grid only at $h=0.6$. We found that in nine of the ten cells there is a bias at which the transverse field improves on the unbiased control beyond the $95\%$ interval (in all ten cells the mean improves). The effect is, therefore, not limited to $n=8$. The bias that maximizes the improvement depends on the size and the noise level, and $h=0.6$ used in the figures is optimal at neither $n=8$ nor $n=12$. We also observe that the ranking of the rules depends on the bias. For $0.2\le h\le0.3$ biased descent matches or exceeds both annealers at every size and every noise level tested up to $f=0.25$ (at $n=14$ we tested this window only at $f=0.1875$). At $h=0.6$ this holds only at $n=8$, whereas descent falls below the quantum schedule at $n=12$ from $f=0.125$ and at $n=14$ from $f=0.1875$ upward. Descent is lower by $0.02$ to $0.05$ up to $f=0.25$ and by $0.08$ at $n=12$, $f=0.3125$. For $h\ge1.2$ every biased rule falls below the unbiased control (at $h=2.4$ all of them return the input unchanged). For each rule taken at its own best bias, we obtained the highest mean overlap with descent in eight of the ten cells. However, the margin of descent over the quantum schedule ($0.01$ to $0.04$) does not exceed the reproducibility of one cell (Appendix). The two exceptions are a difference within the error bars ($n=8$, $f=0.1875$, $-0.010\pm0.017$) and $n=8$ at $f=0.3125$, which lies beyond the retrieval boundary. Therefore, at these sizes and at $L=3$, the linear annealing schedule adds nothing we can detect once the bias is tuned, although this does not hold at every fixed bias. A nonlinear schedule trained on labelled networks does better ($+0.029\pm0.005$ over descent at $n=8$, $L=3$, $f=0.1875$, with the most probable string read out). However, with $100$ shots per probe this falls to $+0.013\pm0.004$, and about $1000$ shots are needed for the full gain. The trained schedule only matches the readout of the nearest stored word, and it has no trained classical schedule to compare with. As expected for a classical pattern-completion task, neither the geometric thermal schedule nor the linear quantum schedule is needed.

At a strict $\geq0.9$ recall threshold the empirical capacity is one pattern over the whole range we can simulate exactly, $n=6$--$14$ [Fig.~\ref{fig:qhnm}(c) shows $n\le10$]. The asymptotic critical load $L_\text{max}\simeq0.14\,n$~\cite{Amit1985}, which tolerates a few wrong spins, gives $0.84$ to $1.96$ across that range, and consequently a capacity of one is compatible with it but does not determine its slope. The success rate at $L=2$ even drifts down with size, from $0.80$ at $n=6$ to $0.72$ at $n=14$. However, at these sizes we cannot resolve the exponent in $L_\text{max}(n)$, and the panel gives only the order of magnitude. The exponential capacity predicted for the biased problem~\cite{Santra2017} concerns its ground state, which a classical solver can also seek, and it is not tested at these sizes.

\emph{Forgetting rate.}---\phantomsection\label{sec:palimpsest}For finite forgetting the network works as a palimpsest memory~\cite{Parisi1986,Nadal1986} of the marginalist type~\cite{Nadal1986}, with the exponential weighting implemented in hardware by the relaxation of the memristor. If pattern $\mu$ is presented at time $t_\mu$, the memory variables at a later time $T$ are
\begin{equation}
\theta_{ij}(T)=\frac{\eta\Gamma_{\rm c}}{\gamma}\big(1-e^{-\gamma\Delta t}\big)\sum_\mu\xi_i^\mu\xi_j^\mu\,e^{-\gamma(T-t_\mu)}.
\label{eq:palimpsest}
\end{equation}
We define the retrievability of this pattern as the fraction of probes (noise fraction $f=0.0625$) for which the annealed readout recovers $\xi^\mu$ exactly,
\begin{equation}
R(T-t_\mu)=\Pr\big[m(\xi^\mu,s)=1\big].
\label{eq:retrievability}
\end{equation}
We streamed $L=20$ patterns through an $n=8$ network at $\gamma=0.3$ [Fig.~\ref{fig:qhnm}(d)] and found that the most recent pattern is recalled in $93\%$ of probes and the next one in $51\%$, whereas the older patterns are lost. The retrievable window of one to two patterns is shorter than the kernel window $1/\gamma\approx3$. From a fit of $\ln R$ against age over ages $0$ to $4$ we obtain a rate of $0.75$ per pattern (bootstrap $95\%$ interval from $0.60$ to $1.07$), which is higher than the kernel rate $0.3$ because exact recovery requires every spin of the pattern. When we stream patterns from the same ensemble at $\gamma\to0$ and the same $\eta$, the most recent one remains retrievable in only $6\%$ of probes, which is the catastrophic interference of a network far beyond capacity.

\subsection{Benchmark on optical quantum states}
\label{sec:qsdl}

To test the network on data, we also ran it on denoising optical quantum states from their Wigner functions. The dataset and the noise model are those of Ref.~\cite{Grzywa2026}, in the version of commit a110562 of Ref.~\cite{QsdlRepo}, and we use them unmodified. The states are drawn from seven families (Fock, coherent, vacuum, thermal, cat, Gottesman--Kitaev--Preskill and binomial) and each state passes, in this order, through three channels drawn independently. These are pure photon loss with probability $0.7$ (transmissivity uniform between $0.6$ and $0.95$), dephasing $\rho_{kl}\to\rho_{kl}\,e^{-\gamma_{\rm d}(k-l)^2/2}$ in the Fock basis with probability $0.3$ ($\gamma_{\rm d}$ uniform between $0.01$ and $0.2$), and a thermal admixture $(1-p_{\rm th})\rho+p_{\rm th}\rho_{\rm th}$ with probability $0.3$ ($p_{\rm th}$ uniform between $0.05$ and $0.3$, mean photon number of the thermal state $\rho_{\rm th}$ uniform between $0.1$ and $1$). The Wigner image of the resulting state is then read out with additive Gaussian detector noise, whose signal-to-noise ratio, relative to the root-mean-square value of the image, is drawn uniformly between $10$ and $30$~dB. In this section fidelity is the root fidelity $\mathrm{Tr}\sqrt{\sqrt{\rho}\,\sigma\sqrt{\rho}}$ of the clean state $\rho$ and the returned state $\sigma$ (not the squared overlap used in the rest of the paper). Returning the state after the channel gives fidelity $0.832$. We use the network as a codebook, i.e., we store binarized noisy images of prototype states by the same Hebbian rule, coarse-grain a noisy Wigner image, project it through a fixed random scattering layer, binarize it, and map the retrieved configuration back to the prototype it matches. For $280$ prototypes we obtained $0.974$ with the best entry of the codebook and $0.880$ with a single-pass correlator readout (this readout is what one pass through the Hebbian matrix computes). However, the values obtained with the attractor dynamics are $0.468$ for two-body descent and $0.388$ for three-body descent (Sec.~\ref{sec:dense}), i.e., far below the $0.832$ found if the state is left unchanged. In our opinion, this failure is caused by the correlations between the prototypes and not by capacity or noise. The average overlap of the prototypes is $0.254$, and for the closest pair it is as large as $0.996$, while for random patterns of the same length we find $0.025$. At the correlations of the prototypes the basins of attraction merge. We also checked that the attractor is not restored if fewer prototypes are stored, down to the $35$ that we tested ($0.44$ and $0.41$). For comparison, a codebook whose best entry reaches $0.93$ needs of order $140$ prototypes. We conclude that a Hopfield attractor is not suited to a task with intrinsically correlated classes.

\section{Coherent recall of superposed inputs}
\label{sec:coherent}

In the retrieval of Sec.~\ref{sec:retrieval} both the input and the answer are classical. In this retrieval the corrupted pattern enters as a longitudinal bias, i.e., as a parameter of the Hamiltonian, which means that there is no input state that could be superposed. The readout is a projection onto the computational basis. Here we test the network without these two restrictions, i.e., we prepare the register in a superposition of two corrupted stored patterns and propagate it unitarily under the linear schedule of Sec.~\ref{sec:retrieval} (Appendix) with $h_i=0$ and no readout. The best of $20$ combinations of the transverse-field amplitude and the annealing time returns the target superposition with fidelity $0.053$ (one draw of three random patterns at $n=8$, one flipped spin in each input, on different sites, mean over the $20$ combinations $0.014$). If we remove the couplings, the same scan gives $0.051$, and we conclude that this number reflects the transverse field. On such different errors the decoder below also stops at $1/2$. A unitary on the register alone cannot contract. For words at distance above two, any such unitary is therefore limited to $1/n$ on average over the $n$ single-bit errors, each acting on both branches.

We therefore define the task of a memory of superposed inputs, in which the memory knows $K$ stored words, i.e., fixed configurations. The memory receives one copy of $\ket{\psi}=(\ket{a}+e^{i\varphi}\ket{b})/\sqrt2$, possibly corrupted, with the pair of words $(a,b)$ and the phase $\varphi$ drawn uniformly and unknown to it, and it has to return $\ket{\psi}$. Distinct words are orthogonal kets, whatever the overlap of their $\pm1$ images. As a result, a readout that returns one configuration, or a mixture of configurations, has fidelity $F=\bra{\psi}\rho_\text{out}\ket{\psi}$ with the target at most $1/2$, where $\rho_\text{out}$ is the returned state, which covers every rule in Fig.~\ref{fig:qhnm}(b). For $K\ge3$ no measure-and-prepare memory, i.e., no entanglement-breaking channel, which measures its input and prepares a state conditioned on the outcome~\cite{Horodecki2003EB}, does better, and the classical register of Part~I is such a memory. On the span of the words such a channel can be taken to have rank-one elements of a positive operator-valued measure (POVM) $w_k\ket{m_k}\bra{m_k}$ with unit vectors $\ket{m_k}$, $\sum_kw_k=K$, and prepared vectors $\ket{g_k}$ of norm at most one. For one element, with $x_c=|\braket{c|m}|^2$ and $y_c=|\braket{c|g}|^2$ for each stored word $c$, averaging over $\varphi$ (denoted $\mathbb{E}_\varphi$), bounding the remaining cosine by one and using $2\sqrt{x_ax_by_ay_b}\le x_ay_b+x_by_a$ gives
\begin{equation}
\mathbb{E}_\varphi|\braket{m|\psi}|^2|\braket{g|\psi}|^2\le\tfrac14\big(x_ay_a+x_by_b+2x_ay_b+2x_by_a\big).
\label{eq:mp_bound}
\end{equation}
Summing over the $K(K-1)/2$ pairs, with $\sum_cx_c=1$, $\sum_cy_c\le1$ and $S=\sum_cx_cy_c\le1$, gives at most
\begin{equation}
\frac{(K-3)S+2}{4}\le\frac{K-1}{4}\quad\text{for }K\ge3.
\label{eq:mp_sum}
\end{equation} After weighting by $\sum_kw_k=K$ and dividing by the number of pairs, we find that the mean fidelity is at most $1/2$, and this value is reached by returning one word. Noise that acts before the measurement and does not depend on the pair or the phase only changes the POVM, and consequently the limit also holds for such corrupted inputs. If the noise reveals the pair, we have instead the higher known-pair limit, i.e., $3/4$ at unknown phase~\cite{Derka1998,Bruss2000} and $2/3$ over all superpositions of the pair~\cite{Massar1995}, while a state known in advance needs no memory. At a known pair and unknown phase the input is an equatorial qubit, for which the best classical copying, which measures the input and prepares the detected state, also reaches $3/4$~\cite{Bartkiewicz2014,Bartkiewicz2017}. This limit is the accessible fidelity of the ensemble~\cite{FuchsSasaki2003}, i.e., the largest mean fidelity that a measure-and-prepare channel reaches on it. Note that benchmarks of this kind are used for quantum storage~\cite{Calsamiglia2009}. If a recall is to exceed this limit, it has to be a map on states.

Maps of this kind are well known from quantum error correction~\cite{NielsenChuang2010}. For a three-qubit repetition code it has been described in dissipative form~\cite{LabayMora2025}, and in Ref.~\cite{Bondarenko2020} quantum autoencoders learned it numerically for Greenberger--Horne--Zeilinger (GHZ) states. If the stored words are chosen in advance for the memory, we refer to them as \emph{codewords}. The recall writes to an auxiliary register (the \emph{ancilla}), the final content of which is referred to as the \emph{syndrome}. We let the recall act as
\begin{equation}
\ket{x}\ket{0}\mapsto\ket{\mathcal{D}(x)}\ket{x\oplus\mathcal{D}(x)},
\label{eq:decoder}
\end{equation}
where $\mathcal{D}$ returns the stored word nearest to $x$ and $\oplus$ is bitwise addition modulo 2, which means that the error is written to the ancilla. It is easy to see that the inputs $\ket{x}\ket{0}$ are mapped to orthonormal outputs, as $x$ is determined by the pair $(\mathcal{D}(x),x\oplus\mathcal{D}(x))$. This map is an isometry, which can be extended to a unitary on the joint register. If we look at the data alone, the map contracts the corrupted inputs, because many of them reach one stored word, whereas the information needed to distinguish them goes to the ancilla. When both branches have the same error, the ancilla reveals nothing about the stored word. With the ancilla reset the decoder is itself a dissipative channel on the data. It should be stressed that the obstacle is the selection of one configuration, and that dissipation itself is not an obstacle. As the map is linear, it maps a superposition of corrupted words to the superposition of their corrected versions.

For $x=c\oplus e$ the ancilla holds $e$ whenever $\mathcal{D}(x)=c$, for any set of stored words, linear or not. Since coherence between two stored words survives only if the ancilla does not reveal which word was stored, the bit-flip error must be the same on both branches. This holds for every Kraus operator $\prod_i(\hat\sigma_x^i)^{e_i}$ of a bit-flip channel acting on the register after the superposition is formed. If the weight of the error is at most the decoding radius $t_\text{dec}=\lfloor(d-1)/2\rfloor$, where $d$ is the minimum Hamming distance between the stored words, then the error is guaranteed to lie in the decoding region of both words. A stochastic mixture of such errors therefore leaves $F=1$. Different errors on the two branches require a corruption that depends on the stored configuration, e.g., a faulty controlled gate, or one that acts on the branches before they are superposed. Unfortunately, this is the bit-flip half of a Calderbank--Shor--Steane (CSS) code~\cite{CalderbankShor1996} and, as a consequence, phase errors escape the decoder. A single $\hat\sigma_z$ on a site where the two words differ gives $F=0$. Dephasing with probability $q$ per qubit gives $F_q$ for words at distance $d_{12}$, and since a photon lost on such a site reveals the branch, loss with transmissivity $\eta_{\rm t}$ without postselection on photon number gives at best $F_{\eta}$, where
\begin{equation}
F_q=\frac{1+(1-2q)^{d_{12}}}{2},\qquad F_{\eta}=\frac{1+\eta_{\rm t}^{d_{12}}}{2}.
\label{eq:phase_loss}
\end{equation} A larger distance thus makes the phase more fragile. We checked the bit-flip condition on four words of length five with $d=3$, which are a coset of a linear code, and on three random words of length nine with $d\ge3$, which form no coset. The fidelity of the reduced data state with the target is $1.000$ for every error of weight at most one and lies between zero and one for larger errors, whereas for branches carrying different single-bit errors we obtain $0.500$ on the coset. It follows that linearity of the code is needed only to compute the syndrome efficiently. For a linear code with parity-check matrix $H$ the syndrome of $x=c\oplus e$ is
\begin{equation}
Hx=Hc\oplus He=He,
\label{eq:syndrome}
\end{equation}
because $Hc=0$ for every codeword. Thus, $Hx$ fixes the coset of the error without a search over the stored words, whereas a decoder for random patterns compares the input with the stored words. Nevertheless, the coset leader still has to be decoded, which is hard in general~\cite{Berlekamp1978}. As an example, we take the first-order Reed--Muller code $\mathrm{RM}(1,m)$ of length $n=2^m$, which is linear and whose $\pm1$ images are the rows of a Hadamard matrix and their complements. An $n=64$ register holds $128$ words for the decoder ($64$ orthogonal rows and their complements), where this count follows from the code and is not a Hebbian capacity, which we treat in Sec.~\ref{sec:codes}. The minimum distance is $d=n/2$, which allows the decoder to correct $t_\text{dec}=15$ flips, i.e., $23\%$ of the word. The syndrome occupies $57$ ancilla qubits and is extracted by $274$ controlled-NOT gates (one per nonzero entry of $H$ in systematic form). It should be noted that decoding it, e.g., by a fast Hadamard transform, is not included in this count.

We tested the recall on $\mathrm{RM}(1,4)$ with all $32$ codewords stored in an $n=16$ register. We drew $500$ times a pair, a relative phase and a bit-flip error of weight one to three. Over these draws the fidelity with the target superposition was $1.000$, i.e., twice the limit of any measure-and-prepare memory. Both branches were restored in $100\%$ of cases and the syndrome agreed in $100\%$ of them. However, if the two corrupted copies are prepared separately with different errors, then the fidelity drops to exactly $1/2$. In this case the syndromes on the branches are different and the ancilla records which branch was corrupted. Note that the decoder itself is a standard one and does not use the memristor. Without errors, the superposition is held with unit fidelity by any coherent register. Therefore, the limit separates coherent storage from measure-and-prepare storage, whereas the tolerance to bit flips is added by the decoder.

To extract the syndrome one needs a network of controlled-NOT gates. The gate of Part~I can serve this purpose, as it is an Ising bond up to local rotations (Sec.~\ref{sec:cz_ising}), but its two fixed loops couple only consecutive photons. Loops of selectable length would thus be required. In addition, the ancilla has to remain coherent with the data register while the syndrome is extracted. In Part~I we tested this requirement only in the two-photon case with an idle gate.

For a linear code $C$ the $p$-body Hebbian sum over all codewords, $\sum_{c\in C}\prod_{i\in\mathcal{T}}\xi_i^c$, equals $|C|$ on the sets $\mathcal{T}$ of $p$ sites that form a parity check and vanishes otherwise. Thus, the resulting energy is a sum of parity checks, as in the energies of Ref.~\cite{Bruck1989}. It should be stressed that we do not claim that the feedback law stores a code, which remains open. This would need signed $p$-fold coincidences (the two- and three-body sums vanish on $\mathrm{RM}(1,m)$) and equal weights, which forgetting breaks. It would also require a set of weight-$p$ checks that generates all checks. In addition, single-spin-flip descent on this energy does not in general reach the decoding radius.

The $p$-body sum is nonzero only on the words of weight $p$ of the dual code $C^\perp$. The order $p$ is thus set by the lightest words of this code. For the simplex code of length $n=2^m-1$, whose $n+1$ codewords lie at the mutual distance $(n+1)/2$~\cite{MacWilliams1977}, the dual is the Hamming code. It has no words of weight two and is generated by its $n(n-1)/6$ words of weight three, so that the two-body sums vanish and the three-body couplings of Sec.~\ref{sec:dense} are sufficient. The energy reaches its minimum only when all these checks are satisfied, i.e., exactly on the codewords (we also enumerated all configurations for $n=7$ and $15$), and every superposition of two codewords is then a ground state. Note that for this family greedy single-spin-flip descent does reach the decoding radius $t_\text{dec}=(n-3)/4$: at $n=7$, $15$ and $31$ it returned the codeword in all $1000$ draws for each error weight up to $t_\text{dec}$. Unfortunately, forgetting lifts the degeneracy of the codewords, which is why the construction needs equal weights. For $n=15$ and one pass over the codewords with the weight reduced by $e^{-\gamma\Delta t}$ per word, the energies of the codewords spread by $12\%$ at $\gamma\Delta t=0.01$. The codewords remain the $16$ lowest states up to $\gamma\Delta t=0.03$, but not at $0.05$, and the spread gives the two branches of a stored superposition different dynamical phases. The learning step would also need one memory variable per triple driven by the sign of triple coincidences, which the device of Part~I does not provide, and the coherent recall still requires the decoder described above.

\section{Coherent-loop realization}
\label{sec:loop}

This section describes a coherent realization of the couplings used in Secs.~\ref{sec:hopfield} and~\ref{sec:coherent}. We first show that a controlled-phase gate between photons is an Ising bond (Sec.~\ref{sec:cz_ising}). In Sec.~\ref{sec:dense} we extend it to operator-valued weights and three-body couplings, and the capacity of codewords is finally compared with that of higher-order couplings in Sec.~\ref{sec:codes}.

\subsection{Controlled-phase gate as an Ising bond}
\label{sec:cz_ising}

Part~I introduced a delay-loop architecture with two fixed loops, in which consecutive photons interact through linear-optical controlled-phase gates, and here we assume loops of selectable length. Since a controlled-phase (CZ) gate is an Ising bond, this extension supplies the bonds of the decoder of Sec.~\ref{sec:coherent} and can also run the schedule of Sec.~\ref{sec:retrieval} coherently, although for this schedule we show no gain. Writing
\begin{align}
\mathrm{CZ}_{ij}(\phi)&=\exp\!\big(i\phi\,\ket{11}\!\bra{11}\big), \nonumber\\
\ket{11}\!\bra{11}&=\tfrac14\big(\openone-\hat\sigma_z^i-\hat\sigma_z^j+\hat\sigma_z^i\hat\sigma_z^j\big),
\label{eq:cz_def}
\end{align}
where $\openone$ is the identity operator, we obtain the exact decomposition
\begin{equation}
\mathrm{CZ}_{ij}(\phi)=e^{i\phi/4}\,
\underbrace{e^{\,i\frac{\phi}{4}\hat\sigma_z^i\hat\sigma_z^j}}_{\text{Hopfield bond}}\;
\underbrace{e^{-i\frac{\phi}{4}\hat\sigma_z^i}\,e^{-i\frac{\phi}{4}\hat\sigma_z^j}}_{\text{local fields}}.
\label{eq:cz_ising}
\end{equation}
A round of tunable~\cite{Lemr2015,Lemr2015JOpt} gates over all pairs is thus one Trotter step of duration $\delta t$ of $e^{-i\delta t\,\hat H}$ for the problem Hamiltonian of Eq.~\eqref{eq:qhn_hamiltonian}, where the weight $J_{ij}=\phi_{ij}/(4\delta t)$ is set by the gate phase. As the length of the loop selects which pairs are coupled, the connectivity is time-multiplexed by the delay loops, and the whole weight matrix is applied sequentially by one physical gate. For the transverse field we propose variable-retardance wave plates at $45^\circ$, since a fixed half-wave plate is a reflection and cannot supply a continuously ramped rotation. By lowering $\Omega$ from its initial value $\Omega_0$ to zero over $M$ round trips, we realize the schedule
\begin{align}
U=\prod_{k=1}^{M}&\Big[\prod_{i<j}\mathrm{CZ}_{ij}(4\lambda_kJ_{ij}\delta t)\Big]
\Big[\prod_i e^{\,i \lambda_k\big(h_i+\sum_{j\neq i}J_{ij}\big)\delta t\,\hat\sigma_z^i}\Big] \nonumber\\
&\times\Big[\prod_i e^{\,i(1-\lambda_k)\Omega_0\delta t\,\hat\sigma_x^i}\Big],
\label{eq:trotter_anneal}
\end{align}
up to a global phase, with the factor $k=1$ acting first and $\lambda_k=(k-\tfrac12)/M$, where the single-qubit $z$ rotations cancel the byproducts of Eq.~\eqref{eq:cz_ising} and imprint the biases $h_i$. The phases $\phi_{ij}$ are classical settings programmed from the registers $\theta_{ij}$ of Sec.~\ref{sec:hebbian}, with the factors $s_k$ and the sums over $j$ computed classically. In this way, the loop uses the Hebbian weights, although a fully coherent, measurement-free learning rule is still an open problem.

\emph{Experimental requirements.}---\phantomsection\label{sec:requirements}With one physical gate each photon makes $n-1$ passes per step, and for a per-pass transmission $\eta_\text{loop}$ the $n$-photon register survives the $M$ annealing steps with probability $\eta_\text{loop}^{n(n-1)M}$. This probability remains of order one for a per-pass loss below $\sim 1/[n(n-1)M]$, which for the schedule simulated here ($n=8$, $M=120$) is $1.5\times10^{-4}$ per pass, or $6.5\times10^{-4}$~dB. This requirement is stricter by about three orders of magnitude than the $1.19$~dB steering budget of Part~I (three settings, ideal gates, unit efficiency outside the loops, random outcomes for no-clicks). Polarization-independent loss erases a photon and leaves the surviving runs coherent, so that above this threshold only the rate vanishes. However, the dominant experimental obstacle is the nondeterminism of the gates. Postselected linear-optical gates~\cite{Lemr2015} cannot be cascaded over the $Mn(n-1)/2$ bond applications of Eq.~\eqref{eq:trotter_anneal}, and hence the coherent mode requires near-deterministic, heralded gates with feed-forward~\cite{KLM2001}, as do the $274$ controlled-NOT gates of the syndrome in Sec.~\ref{sec:coherent}. Note that Eq.~\eqref{eq:trotter_anneal} reproduces the annealing schedule simulated in the Appendix up to Trotter error, as Eq.~\eqref{eq:trotter_anneal} is a first-order product while the simulation uses symmetric second-order steps. A quantitative account of the Trotter error and an integrated-photonics realization require a separate study.

\subsection{Operator-valued weights and three-body couplings}
\label{sec:dense}

The coherent network adds two ingredients. The first one is a program register in superposition, which coherently superposes distinct weight matrices if the gate phase is set by a \emph{program} photon (this itself requires a controlled-controlled-phase gate). The analysis of the readout protocol, which determines the information that can be extracted per run, is beyond the scope of this paper. The second ingredient is given by higher-order couplings. The three-photon gate of Part~I applies only two-qubit phases, but if we upgrade the loop gate to a three-photon controlled-controlled-phase (CCZ) gate, such couplings are obtained from linear optics and entanglement together with measurement, without a physical optical nonlinearity. Nondeterministic linear-optical gates of this kind have been demonstrated at $\phi=\pi$ on three qubits and at tunable phase on two~\cite{Lanyon2009}. The gate acting on qubits $i$, $j$ and $k$ obeys
\begin{align}
\mathrm{CCZ}_{ijk}(\phi)&=\exp\!\big(i\phi\ket{111}\!\bra{111}\big), \nonumber\\
\ket{111}\!\bra{111}&=\tfrac18\Big(\openone-\sum_{a}\hat\sigma_z^a+\sum_{a<b}\hat\sigma_z^a\hat\sigma_z^b-\hat\sigma_z^i\hat\sigma_z^j\hat\sigma_z^k\Big),
\label{eq:ccz}
\end{align}
where $a,b\in\{i,j,k\}$, and therefore contains a term $-J_{ijk}\hat\sigma_z^i\hat\sigma_z^j\hat\sigma_z^k$ of $\hat H$ with $J_{ijk}=-\phi_{ijk}/(8\delta t)$. Its two- and one-body terms are cancelled by $\mathrm{CZ}_{ab}(-\phi_{ijk}/2)$ on each pair and $e^{-i\phi_{ijk}\hat\sigma_z^a/8}$ on each qubit. If these terms are left in place, they lower the exact-recovery capacity of Sec.~\ref{sec:codes} to $33$ on average at $n=64$ ($19$ to $49$ over $12$ seeds), which is below the $n$ words of an orthogonal code. The pattern-dependent $J_{ijk}\propto\sum_\mu\xi_i^\mu\xi_j^\mu\xi_k^\mu$ requires the phase to be set per triple, as Sec.~\ref{sec:cz_ising} assumes for $\phi_{ij}$, and the drawback of this construction is that we derive no feedback law for it. Such couplings define higher-order Hopfield networks~\cite{Gardner1987,AbbottArian1987,KrotovHopfield2016,Demircigil2017}, whose capacity grows as $n^{p-1}$ for $p$-body couplings, divided by $\ln n$ for exact recall. Zanfardino \emph{et al.} showed that indistinguishable photons in a linear interferometer already produce statistics governed by a $p$-body Hopfield Hamiltonian, and that the four-body model has a transition from retrieval to a spin-glass phase~\cite{Zanfardino2026}, which was then observed on programmable interferometers~\cite{Giordani2026}. As a result, many-body Ising terms do not require an optical nonlinearity, but in their scheme $p$ is even and the couplings are permanents of the scattering matrix (they are not per-bond weights), while our weights $J_{ijk}$ are set per triple with $p=3$.

\begin{figure}[tb]
\centering
\includegraphics[width=\columnwidth]{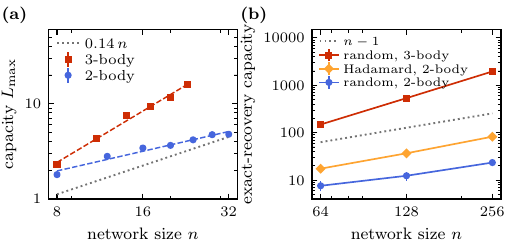}
\caption{Capacity by zero-temperature descent. (a)~Largest pattern set retrievable at $\ge0.9$ success from inputs with $\max(1,\mathrm{round}(0.08\,n))$ flipped spins versus $n$ for the two-body (CZ) and three-body (CCZ) energy functions. Each point is the mean over $40$ realizations ($60$ trials per pattern), with error bars giving the standard error. Dashed lines are log--log fits, and the dotted line is $0.14\,n$. (b)~Exact-recovery capacity under the protocol of Sec.~\ref{sec:codes} for random patterns with three-body couplings, random subsets of Hadamard codewords with two-body couplings, and random patterns with two-body couplings ($12$ seeds, error bars giving the standard deviation), and the limit $n-1$ of the codewords (dotted).}
\label{fig:dense}
\end{figure}

In Fig.~\ref{fig:dense}(a) we show classical zero-temperature descent on the energies
\begin{equation}
E_p(s)=-\sum_\mu\Big(\sum_i\xi_i^\mu s_i\Big)^p,\qquad p=2,3,
\label{eq:dense_energy}
\end{equation}
which checks the known capacity and is not a simulation of the quantum network. For $p=3$ the terms with repeated indices add a one-body field to the couplings $J_{ijk}$, available from $z$ rotations, and without this field the capacity is higher. The log--log fits have slopes $0.69\,[0.63,0.74]$ for two-body and $1.79\,[1.68,1.89]$ for three-body couplings (brackets are bootstrap $95\%$ intervals). These values exceed the capacity of one pattern of Sec.~\ref{sec:retrieval} because in this panel we use descent and average the largest retrievable set of each realization. In Fig.~\ref{fig:qhnm}(c), on the other hand, we use the anneal and the threshold is applied to the success averaged over networks. Note that the noise level is not the cause of this difference, as at the shared size $n=8$ both protocols flip one spin (in Sec.~\ref{sec:codes} we flip $\lfloor0.15\,n\rfloor$ spins, so the values there are not on the scale of this panel). Recall counts when at least $99\%$ of the spins are correct, which at these sizes means every spin, with capacity $n/(2\ln n)$ for $p=2$ and $n^2/(6\ln n)$ for $p=3$~\cite{KrotovHopfield2016}. Their local slopes on these grids are $0.63$ and $1.61$, and therefore neither fit gives an asymptotic exponent. Although this range alone does not exclude a linear law with an offset, superlinear growth follows from the larger registers of Sec.~\ref{sec:codes}.

\subsection{Codewords versus higher-order couplings}
\label{sec:codes}

In Fig.~\ref{fig:dense}(b) we compare what is stored with the order of the couplings. Random patterns are far from the best case for a Hebbian rule, although the correlated patterns of Sec.~\ref{sec:qsdl} are worse. The crosstalk on a stored word grows as $\sqrt{L/n}$. Orthogonal codewords, e.g., the rows of a Hadamard matrix, have zero mutual overlap~\cite{MacWilliams1977}. It follows that up to $n-1$ of them are fixed points~\cite{Hertz1991}. Note that these fixed points do not imply retrieval. With $n-1$ rows the couplings are $J_{ij}=-1/n$ for all pairs, so that every balanced configuration is a fixed point, and descent from a corrupted probe returns no word exactly. Similarly, we do not test the network by identifying the nearest stored word, because the corrupted probe itself passes this test far above the numbers below. We therefore take the capacity, which we find by bisection, as the largest number of stored patterns for which descent returns the intended pattern exactly in $90\%$ of probes with $\lfloor0.15\,n\rfloor$ flipped spins. For random patterns we obtain the two-body capacity $7.6\pm1.8$, $12.3\pm2.7$ and $23.4\pm2.7$ at $n=64$, $128$ and $256$ (standard deviation over $12$ seeds), which is of the order of $n/(2\ln n)$. Random subsets of Hadamard rows give $17.4\pm1.4$, $36.8\pm1.3$ and $82.3\pm2.8$, i.e., a factor of $2.3$ to $3.5$ which is available from ordinary controlled-phase gates and a better choice of stored words. The decoder of Sec.~\ref{sec:coherent} requires the complete code, which this classical attractor storage of a subset of the words does not provide. Musa \emph{et al.} implement a four-body term through second-harmonic generation in a periodically poled lithium niobate crystal and test it on Hadamard patterns of $100$ spins, with a capacity of $5$ for two-body and at least $50$ with added four-body couplings~\cite{Musa2026}. Their term requires a physical nonlinearity, which the three-photon gate replaces by measurement.

However, orthogonality limits a set of codewords to $n$ words, whereas the three-body capacity continues past this limit. For random patterns with three-body couplings we obtain $149\pm12$, $535\pm37$ and $1953\pm158$ at the same sizes, i.e., a factor of $2.4$, $4.2$ and $7.7$ beyond $n-1$. A log--log fit gives $L_\text{max}\propto n^{1.86}$ with a bootstrap $95\%$ interval $[1.82,1.90]$, of the order of the $n^2/(6\ln n)$ expected for exact recovery~\cite{KrotovHopfield2016} ($164$, $563$ and $1970$). Higher-order couplings pass the orthogonality limit (which is itself not a retrieval capacity), but we find that they store no more bits per coupling, i.e., $nL_\text{max}$ bits per $\binom{n}{p}$ $p$-body couplings ($0.18$ at $n=256$ for both orders).

Codewords also fix the error-correcting distance. Hadamard codewords are equidistant at $d=n/2$, while the minimum distance of a random set fluctuates and is lower. At $n=256$ with $32$ stored words the Hadamard set has $d=128$ and a random set has median $104$ over $10^5$ draws (interquartile range $102$--$106$). This corresponds to $t_\text{dec}=63$ against $t_\text{dec}=51$, which is a modest gain compared with the efficient syndrome of Sec.~\ref{sec:coherent}.

\section{Conclusions}
\label{sec:conclusions}

In this paper, we have studied the computational capabilities of networks of the single-photon memristors of Part~I. A single device cannot correct the error of the photon it transmits. However, a register of pairwise-coupled devices learns the Hebbian matrix from the feedback law, and a network programmed with this matrix completes corrupted patterns through attractor dynamics. Quantum-annealing retrieval appears to beat greedy descent only because it alone receives the corrupted input as a longitudinal bias. Once that bias is shared and set in the window $0.2\le h\le0.3$, zero-temperature descent matches or exceeds the annealers at $n=8$ and $12$, and at $n=14$ at the one noise level tested. The annealers give a higher recall than descent at $h=0.6$ (by $0.02$ to $0.05$ up to $f=0.25$), which holds at $n=12$ from $f=0.125$ and at $n=14$ from $f=0.1875$. At these sizes associative recall of classical patterns yields no detectable quantum advantage and, with the bias tuned, requires no linear annealer. On correlated data (here optical states) attractor dynamics falls below leaving the input unchanged. With finite forgetting we obtain a palimpsest, the window of which is tuned by one physical parameter. Finally, in the coherent loop the Ising bonds follow from an exact decomposition of the gate, and three-body couplings are supplied by a three-photon controlled-controlled-phase gate without a physical optical nonlinearity.

We also defined a memory task in which the register receives an equal superposition of two of $K$ stored words, with the pair and the relative phase unknown. For $K\ge3$ no measure-and-prepare memory exceeds mean fidelity $1/2$. However, the coherent decoder gives unit fidelity under bit-flip noise within the decoding radius, which acts alike on both branches. Every other gain reported here is matched by classical means (a better encoding, a higher-order network or biased descent), and the trained schedule only equals the readout of the nearest stored word. It should be stressed that this gain comes from the coherent decoder, which is standard syndrome decoding and does not use the memristor, and that it does not extend to phase errors or photon loss. Unfortunately, we have not shown that the feedback law stores codewords, and the unitary evolution under its couplings gives fidelity $0.053$ in a 20-point scan on one draw, against $0.051$ with the couplings removed.

Note that coherent memory is already used as a figure of merit in quantum cryptography, where it lets an eavesdropper defer the measurement until the basis reconciliation~\cite{Bartkiewicz2013,Bocquet2012}. This figure of merit also appears in the counterfeiting of quantum money, for which an attack that measures and prepares stays at the fidelity of classical copying, while optimal cloning exceeds this fidelity~\cite{Bartkiewicz2017,Jirakova2019}. In learning theory, exponential separations are known between protocols that do and do not retain quantum memory between experiments~\cite{Huang2022,Chen2021}. Although these parallels are only qualitative, in both cases the memory has to survive coherently until the information it must be combined with arrives, which is the property that Part~I proposes to certify.

\section*{Acknowledgments}
We thank F.~Grzywa for the optical quantum-state dataset and the noise model used in Sec.~\ref{sec:qsdl}, which were developed within the project \textit{Deep learning for the classification and reconstruction of optical quantum states}, carried out under the Excellence Initiative--Research University (IDUB) programme at Adam Mickiewicz University. We acknowledge funding from the European High Performance Computing Joint Undertaking (EuroHPC JU) under Horizon Europe Grant Agreement No.~101194322 (QEC4QEA), co-funded by the Polish National Centre for Research and Development (NCBiR) under decision No.~DWM/EuroHPC/2023/429/2025. The text was drafted and revised with the AI assistant Claude Code (Anthropic), and the authors take full responsibility for the article.

K.B. conceived the study, derived the analytical results, developed the simulation code and wrote the manuscript with the AI assistance stated in the Appendix and in the Acknowledgments, and supervised the work. P.T. and A.M. contributed to the physical model of the device and to the interpretation of the results. All authors discussed the results and revised the manuscript.

The authors declare no competing interests.

\section*{Data availability}
All simulation code and all data behind the figures and the numbers quoted in the text are openly available in the repository of Ref.~\cite{Repo2026}. This repository contains the deterministic scripts and the cached results of the expensive runs, so that every figure can be regenerated without rerunning the simulations. The optical quantum-state dataset and noise model used in Sec.~\ref{sec:qsdl} are those of Ref.~\cite{QsdlRepo}. No experimental data were generated in this work.

\appendix

\section{Numerical methods}
\label{app:methods}

\textit{Learning.} We stream random $\pm1$ patterns through Eq.~\eqref{eq:theta_dynamics} with $n=8$, $L=5$, $\eta=1/n$, $\gamma=0.01$, $\Gamma_{\rm c}=1$, $\Delta t=1$. The deterministic drive gives $r=0.9999$ with Eq.~\eqref{eq:hebbian}, limited by the weak forgetting, and in Fig.~\ref{fig:qhnm}(a) we report the Poisson-sampled drive ($200$ streams per point).

In the \textit{retrieval} we apply $\tilde\xi$ as a local bias $h_i=0.6\,\tilde\xi_i$ and let the register relax toward the ground state of
\begin{equation}
\hat H(\lambda)=-(1-\lambda)\,\Omega_0\sum_i\hat\sigma_x^i+\lambda\Big[-\!\sum_{i<j}J_{ij}\hat\sigma_z^i\hat\sigma_z^j-\sum_i h_i\hat\sigma_z^i\Big]
\label{eq:anneal_schedule}
\end{equation}
by real-time Schr\"odinger evolution of the exact $2^n$-dimensional state from $\ket{+}^{\otimes n}$ ($\Omega_0=1$, total annealing time $T_\text{a}=12$, $M=120$ symmetric second-order Trotter steps, midpoint schedule linear in $\lambda\in[0,1]$). Then, we read out the most probable basis string (an infinite-shot idealization). We count a retrieval as successful when $m>0.999$. As the unbiased control we use greedy single-spin-flip descent on the couplings alone, which is started from $\tilde\xi$. The biased control is the same descent applied to the input-biased problem. To compare the quantum schedule with a classical heuristic, we have also implemented simulated annealing based on the Metropolis dynamics for the same problem. In our simulations the temperature has been decreased geometrically. Its initial value was $T_0=1.5\,\bar h$, where $\bar h$ denotes the median magnitude of the local field, and the final one was $0.02\,\bar h$. Each run consisted of $300$ sweeps and it was followed by a settle at zero temperature. The annealing has been restarted three times and only the best result was kept. As for the uncertainties, the bands shown in our plots have been calculated as $1.96\,\sigma/\sqrt{N}$, where $\sigma$ and $N$ refer to the pattern sets. To be sure that this simple estimate is good enough, we have compared it with percentile bootstrap intervals and the two agree to a median of $2.7\%$. Note that for different disorder draws the result for a single (size, noise, bias) cell is reproducible only to about $\pm0.04$ (at most $0.06$). Hence, we do not infer any mechanism from one cell.

We have also scanned the annealer over its own parameters. For this scan we used the first of the six further draws of $60$ networks of Sec.~\ref{sec:retrieval}, with $100$ probes per network, at $n=8$, $L=3$ and $f=0.1875$. The annealing time took the values $T_\text{a}\in\{6,12,25,50\}$ and the transverse-field amplitude the values $\Omega_0\in\{0.25,0.5,1,2\}$. We found that the recall saturates at $0.747\pm0.015$. This value is reached in the best cell, $(T_\text{a},\Omega_0)=(50,0.5)$, whereas for the setting used in the paper, $(12,1)$, we obtained $0.729\pm0.016$. On the same data biased descent gave $0.754\pm0.015$ and simulated annealing $0.742\pm0.015$. For the unbiased control the recall was $0.643\pm0.022$. We have also compared the best anneal with biased descent network by network and found that it is lower by $0.008\pm0.007$. From simulated annealing it cannot be distinguished, as the difference is $+0.006\pm0.009$. We have repeated the same scan at $f=0.125$ on two further draws. There the best anneal [$T_\text{a}\ge25$, $\Omega_0\le1$] comes to within $0.004$--$0.010$ of biased descent, while at $(12,1)$ the distance is $0.046$--$0.051$. Most of the gap quoted in Sec.~\ref{sec:retrieval} at that noise is, therefore, due to the fixed schedule. Finally, we have tested the Trotterized propagator at a bias far above the coupling scale. In this regime the overlap must be equal to $|n-2\,\mathrm{round}(fn)|/n$. At $n=8$, $12$ and $14$ the propagator reproduces this value up to $h\,\delta t\approx3.3$, where $\delta t=T_\text{a}/M$ is the Trotter step, and it fails at $3.6$. All results reported in this paper lie well below this value.

\textit{Panels of Fig.~\ref{fig:qhnm}.} Panel (b) uses a $17$-point noise grid. For panel (c) we use $60$ networks $\times$ $8$ probes per $L$, and we repeat the same protocol at $n=12$ and $14$. Panel (d) uses $\eta=\gamma=0.30$, $60$ streams $\times$ $12$ probes, $f=0.0625$, and a control at $\gamma=10^{-3}$.

\textit{Capacity.} For Fig.~\ref{fig:dense}(a) we use $2000$ bootstrap resamples of the realizations. In Sec.~\ref{sec:codes} the stored patterns are random or a random subset of the rows of a Sylvester Hadamard matrix without the all-ones row, and in the descent on Eq.~\eqref{eq:dense_energy} we use at most $60$ sweeps. We test each number of patterns with $120$ probes and find the capacity by doubling and bisection with a tolerance of $1\%$ and an upper limit of $16384$ patterns, and the seeds are $100+7s$ with $s=0,\dots,11$.

{\raggedright\textit{Use of AI.} We used the AI assistant Claude Code (Anthropic; models Claude Sonnet 5.5 and Claude Opus 5.5, identifiers claude-sonnet-5-5 and claude-opus-5-5, in the final stage; model versions of earlier stages were not recorded) to write and run the code of the numerical simulations and figures, to derive and check analytical results, including the bound of Sec.~\ref{sec:coherent}, to check the numbers and the bibliography against the code and the sources, and to audit the results by independent reimplementation. We directed this work and checked the results by independent verification with separate programs and by manual consistency checks.

\textit{Scripts} are in the repository of Ref.~\cite{Repo2026}. The network is held in \texttt{sim\_qhnm.py}. We draw all figures from cached data with \texttt{fig\_redesign.py}, and the images of Fig.~\ref{fig:concept} are rendered by \texttt{fig\_concept\_data.py} and \texttt{fig\_concept\_render.py}. We compute the data of Figs.~\ref{fig:qhnm} and~\ref{fig:dense}(a) with \texttt{fixes\_compute.py}, with panel (b) of Fig.~\ref{fig:qhnm} from \texttt{panel\_b\_big.py} and panel (c) from \texttt{make\_paper\_figures.py}. Capacities come from \texttt{capacity\_code.py} and \texttt{capacity\_summary.py}, minimum distances from \texttt{check\_mindist.py}, the unitary scan from \texttt{check2\_quantum\_input.py}, the decoder from \texttt{check4\_qhop.py}, \texttt{check5\_rm\_decoder.py} and \texttt{check\_nonlinear\_code.py}, simplex codes from \texttt{check\_simplex\_codes.py}, the scans from \texttt{scale\_fig2b.py}, \texttt{sweep\_bias.py}, \texttt{sweep\_noise\_bias.py} and \texttt{sweep\_anneal\_tuning.py}, the independent redraws from \texttt{redraw\_panel\_b.py}, further controls from \texttt{check6\_supporting.py}, the Trotter check from \texttt{check\_trotter.py}, the prototype overlaps from \texttt{qsdl\_overlaps.py}, the control at $L=2$ from \texttt{check\_L2\_control.py}, the annealer scan at $f=0.125$ from \texttt{sweep\_anneal\_tuning\_f0125.py}, the palimpsest fit from \texttt{fit\_palimpsest.py} and the trained schedule from \texttt{trained\_schedule.py} and \texttt{trained\_schedule\_test.py}, and the benchmark from \texttt{framework.py}, \texttt{experiment.py} and \texttt{superpos.py}, which additionally require the package of Ref.~\cite{QsdlRepo}.\par}

\end{document}